\documentclass[sn-vancouver,Numbered]{sn-jnl}

\usepackage{graphicx}%
\usepackage{multirow}%
\usepackage{amsmath,amssymb,amsfonts}%
\usepackage{amsthm}%
\usepackage{mathrsfs}%
\usepackage[title]{appendix}%
\usepackage{xcolor}%
\usepackage{textcomp}%
\usepackage{manyfoot}%
\usepackage{booktabs}%
\usepackage{algorithm}%
\usepackage{algorithmicx}%
\usepackage{algpseudocode}%
\usepackage{listings}%
\usepackage{bm,upgreek}
\usepackage{nameref}
\usepackage{soul}

\newcommand\rd{{\rm{d}}}

\DeclareMathOperator{\tr}{tr}

\graphicspath{{Figs/}}

\unnumbered

\begin{document}

\title{A note on thermomechanical coupling effects in the indentation of pseudoelastic shape memory alloys$^\dagger$}

\author*[1]{\fnm{Mohsen} \sur{Rezaee-Hajidehi}}\email{mrezaee@ippt.pan.pl}
\author[1]{\fnm{Mahdi} \sur{Neghabi}}\email{mneghabi@ippt.pan.pl}
\author[1]{\fnm{Stanis{\l}aw} \sur{Stupkiewicz}}\email{sstupkie@ippt.pan.pl}

\affil{\orgdiv{Institute of Fundamental Technological Research (IPPT)}, 
\orgname{Polish Academy of Sciences}, \orgaddress{\street{Pawi\'nskiego 5B}, \city{Warsaw}, \postcode{02-106}, \country{Poland}}}

\abstract{
While macroscopic experiments on polycrystalline shape memory alloys (SMAs) reveal significant thermomechanical coupling effects arising from the latent heat of transformation, the relevance of thermomechanical couplings in indentation tests remains ambiguous. This ambiguity is further emphasized by the rate effects observed in a number of micro/nano-indentation experiments, thus highlighting the need for a more careful investigation of the thermomechanical interactions at such small scales. With this in mind, the present study aims to demonstrate the role of thermomechanical couplings in indentation-induced martensitic transformation in SMAs. To this end, a simple phenomenological model of pseudoelasticity is employed and finite-element simulations are performed to address two key questions. (1) At which spatial and temporal scales do the thermomechanical couplings arising from the latent heat become effective? (2) To what extent do these couplings influence the indentation response? In connection with the latter, our analysis quantifies the maximal thermal effects that emerge during adiabatic indentation and compares them with those of isothermal indentation.
\footnotetext[2]{To appear in \emph{Shape Memory and Superelasticity}}
}

\keywords{Pseudoelasticity; Indentation; NiTi; Latent heat; Thermomechanical interactions}

\maketitle

\section{Introduction}\label{sec-int}
At macroscopic scale, shape memory alloys (SMAs) exhibit a rate-dependent behavior. The rate-dependence is primarily driven by thermomechanical interactions originating from the latent heat of martensitic phase transformation. This is manifested under non-isothermal loading conditions, where the generated latent heat is not completely removed from the specimen due to the limited heat exchange with the surroundings. Accordingly, the resulting temperature rise elevates the transformation stress (in accordance with the Clausius--Clapeyron relation) and thus leads to complex transformation evolutions \citep{shaw1995thermomechanical,zhang2010experimental,xiao2020rate,tsimpoukis2024rate}.

Micro/nano-indentation tests have been extensively used to probe the microstructural features and small-scale mechanical properties of SMAs, e.g., \citep{gall2001instrumented,frick2006stress,pfetzing2013crystallographic,kumar2020assessment}. Within this context, a number of experiments have explored the effect of indentation loading rate on the deformation behavior of the material \citep{amini2011loading,amini2013phase,kumar2020assessment,li2024superelastic}. Notably, in some cases, a rate-dependent behavior has been observed, which has been occasionally attributed to the thermomechanical interactions. It is, however, important to acknowledge that at such small scales, heat conduction is dominant, and the generated latent heat is expected to be rapidly transferred away from the indented region, hence leading to a (nearly) uniform temperature distribution. As a consequence, it can be reasonably inferred that thermomechanical interactions are less likely to be the origin of rate effects in micro/nano-indentation. Nevertheless, the ambiguity surrounding the potential contribution of thermomechanical couplings at small scales underscores the need for further investigation.

The present study aims to elucidate the role of thermomechanical interactions in the indentation-induced martensitic transformation in SMAs. Our investigation is based on a simple phenomenological model of pseudoelasticity and is performed in two steps. First, we employ a thermomechanically-coupled model and assess the significance of thermomechanical coupling effects across various spatial and temporal scales. The goal of this analysis is to address the question of when and how do the transient thermal effects arising from the latent heat of transformation come into play during the indentation of SMAs. Subsequently, we further simplify the model and explore the thermal effects during adiabatic indentation. The goal of this analysis is to provide an upper-bound for the potential thermomechanical coupling effects. 
To reinforce the validity of our findings. we also examine the impact of residual deformation stemming from transformation-induced plasticity on thermomechanical interactions.
The study concludes with a summary of the simulation results and some general remarks. 

\section{Thermal Effects during Indentation of SMAs}\label{sec-thermal}
This section begins with a brief description of the thermomechanically-coupled model of pseudoelasticity, followed by the presentation and discussion of the corresponding simulation results. Subsequently, we employ an extended version of the model that incorporates transformation-induced plasticity and investigate the potential influence of residual deformation on the results.

\subsection{Thermomechanically-Coupled Model of Pseudoelasticity}\label{sec-modelcoup}
A small-strain model of pseudoelasticity is employed. The model can be considered as a simplified non-gradient version of the model developed by \citet{Rezaee2020Gradient}. Formulated within the incremental energy minimization framework, the constitutive description relies on two main elements: the Helmholtz free energy function $\phi$ and the dissipation potential $D$. For simplicity, it is assumed that $\phi$ comprises only the chemical energy $\phi_\text{chem}$ and the elastic strain energy $\phi_\text{el}$. This assumption leads to a flat (neither hardening- nor softening-type) flag-shaped material response. At the same time, the dissipation potential $D$ accounts solely for the dissipation due to martensitic phase transformation, while neglecting dissipation due to martensite reorientation. Following \citep{RezaeeStupkiewicz2018Gradient,Rezaee2020Gradient}, the model incorporates thermomechanical couplings through the chemical energy $\phi_\text{chem}$ and through the internal heat source associated with phase transformation in the heat conduction equation.
The model formulation is presented below. 

In the small-strain setting, the total strain $\bm{\varepsilon}$ is additively decomposed into the elastic and transformation parts,
\begin{equation}
    \bm{\varepsilon}=\bm{\varepsilon}^\text{e}+\bm{\varepsilon}^\text{t}.
\end{equation}
The transformation strain $\bm{\varepsilon}^\text{t}$ is defined in terms of two internal variables, the limit transformation strain, $\bar{\bm{\varepsilon}}^\text{t}$, and the volume fraction of martensite, $\eta$, i.e.,
\begin{equation}\label{eq-const}
    \bm{\varepsilon}^\text{t}= \eta \bar{\bm{\varepsilon}}^\text{t}, \quad \bar{\bm{\varepsilon}}^\text{t} \in \bar{\mathcal{P}} = \{\, \bar{\bm{\varepsilon}}^\text{t}:\, g(\bar{\bm{\varepsilon}}^\text{t})=0 \,\}, \quad 0 \leq \eta \leq 1,
\end{equation}
where the set $\bar{\mathcal{P}}$ of admissible limit transformation strains is represented by the surface $g(\bar{\bm{\varepsilon}}^\text{t})=0$, and the volume fractions $\eta=0$ and $\eta=1$ indicate the pure austenite and pure martensite phases, respectively. It is here assumed that the martensitic transformation is isochoric, implying ${\tr \bar{\bm{\varepsilon}}^\text{t}=0}$, where $\tr$ denotes the trace of a tensor. Additionally, we postulate that the transformation is isotropic and exhibits no tension--compression asymmetry. Consequently, the function $g(\bar{\bm{\varepsilon}}^\text{t})$ depends solely on the second invariant $I_2$ of the limit transformation strain $\bar{\bm{\varepsilon}}^\text{t}$, cf.~\citep{sadjadpour2007micromechanics}, and is defined as
\begin{equation}
    g(\bar{\bm{\varepsilon}}^\text{t})= \sqrt{-I_2}-a, \quad I_2=-\frac{1}{2} \tr (\bar{\bm{\varepsilon}}^\text{t})^2,
    \end{equation}
where $a=\sqrt{3}/2 \epsilon_\text{T}$, with $\epsilon_\text{T}$ representing the maximum attainable transformation strain. It is worth mentioning that the omitted features, namely tension--compression asymmetry and transverse isotropy, are accounted for in the original (finite-strain) model. For further details, see \citep{Rezaee2020Gradient}.

As mentioned above, only the contributions from the chemical energy and the elastic strain energy are considered in the Helmholtz free energy function $\phi$, i.e.,
\begin{equation}    \phi(\bm{\varepsilon},\bar{\bm{\varepsilon}}^\text{t},\eta,T)=\phi_\text{chem}(\eta,T)+\phi_\text{el}(\bm{\varepsilon},\bar{\bm{\varepsilon}}^\text{t},\eta).
\end{equation}
The chemical energy $\phi_\text{chem}$ is given by
\begin{equation}\label{eq-chem}
    \phi_\text{chem}(\eta,T)= \phi_0^\text{a}(T) + \Delta \phi_0 (T) \eta, \quad \Delta \phi_0(T)= \Delta s^* (T - T_\text{t}),
\end{equation}
where $T$ is the temperature, $\phi_0^\text{a}$ is the free energy density of unstressed pure austenite phase, $\Delta s^*$ is the specific entropy difference, and $T_\text{t}$ is the transformation equilibrium temperature. The linear dependence of the chemical energy $\phi_\text{chem}$ on the temperature $T$ through $\Delta \phi_0$ expresses the Clausius--Clapeyron relation and results in a linear dependence of the transformation stress on the temperature \citep{otsuka2005physical}.

The elastic strain energy $\phi_\text{el}$ takes the form
\begin{equation}\label{eq-el}
    \phi_\text{el}(\bm{\varepsilon},\bar{\bm{\varepsilon}}^\text{t},\eta)=\mu \tr(\bm{\varepsilon}^\text{e}_\text{dev})^2+\frac{1}{2} \kappa (\tr \bm{\varepsilon}^\text{e})^2,
\end{equation}
where $\bm{\varepsilon}^\text{e}=\bm{\varepsilon}-\bm{\varepsilon}^\text{t}$ and $\bm{\varepsilon}^\text{e}_\text{dev}$ is its deviatoric part, $\mu$ is the shear modulus and $\kappa$ is the bulk modulus. Although the shear modulus $\mu$ typically varies with the volume fraction $\eta$, due to the distinct elastic properties of the austenite and martensite phases, we assume for simplicity that $\mu$ remains constant during the transformation.

Next, a rate-independent dissipation potential is adopted in the following (incremental) form,
\begin{equation}\label{eq-diss}
    \Delta D(\Delta \eta)=f_\text{c} |\Delta \eta|,
\end{equation}
where $\Delta \eta=\eta-\eta_n$, with $\eta_n$ as the martensite volume fraction related to the previous time-step, and $f_\text{c}$ is the critical driving force for transformation, a parameter that characterizes the width of the hysteresis loop. 

With the Helmholtz free energy $\phi$ and the dissipation potential $\Delta D$ at hand, their global counterparts are constructed, i.e., $\Phi=\int_B \phi \, \rd V$ and $\Delta \mathcal{D}=\int_B \Delta D \, \rd V$. A global incremental potential $\Pi$ is then formulated as the sum of the global incremental energy $\Delta \Phi$, the global dissipation potential $\Delta \mathcal{D}$, and the potential of external loads $\Delta \Omega$ (left unspecified). Subsequently, at a given temperature $T$, the solution of the problem in terms of the displacement $\bm{u}$, the limit transformation strain $\bar{\bm{\varepsilon}}^\text{t}$ and the martensite volume fraction $\eta$ is sought through the minimization of the incremental potential $\Pi$. This is expressed as
\begin{equation}\label{eq-min}
    \Pi=\Delta \Phi + \Delta \mathcal{D} + \Delta \Omega \; \rightarrow \; \min_{\substack{\bm{u},\bar{\bm{\varepsilon}}^\text{t},\eta}} \quad (\text{at a given }T).
\end{equation}
Note that the minimization problem~\eqref{eq-min} is subject to equality and inequality constraints on the internal variables $\bar{\bm{\varepsilon}}^\text{t}$ and $\eta$, see Eq.~\eqref{eq-const}$_{2,3}$.

To complete the model, the temperature $T$ is determined by solving the heat conduction equation. A thermomechanically-coupled model is achieved by formulating the internal heat source $\dot{R}$ to account for the two primary heat contributors, namely the latent heat of transformation and the heat produced by mechanical dissipation. Thus, $\dot{R}$ takes the form
\begin{equation}
    \dot{R}=\Delta s^\ast T \dot{\eta}+f_\text{c}|\dot{\eta}|,
\end{equation}
where $\dot{\eta}=\Delta \eta/\Delta t$ in the incremental setting. The (isotropic) heat conduction equation is then given by
\begin{equation}\label{eq-heat}
    \varrho_0 c \dot{T} + \nabla \cdot \bm{q} = \dot{R}, \quad \bm{q}=-k \nabla T,
\end{equation}
where $\varrho_0 c$ is the specific heat capacity, $k$ is the heat conduction coefficient, both assumed identical for austenite and martensite phases, and $\bm{q}$ is the heat flux. It is important to note that, in this study, the heat exchange between the material and the indenter as well as the heat convection with the ambient are neglected. Incorporating these heat exchange processes would reduce the thermal effects, hence the present analysis offers an upper bound estimation of the thermomechanical interactions.

The minimization problem \eqref{eq-min} can be reformulated as a global--local minimization problem, where the limit transformation strain $\bar{\bm{\varepsilon}}^\text{t}$ and the martensite volume fraction $\eta$ are the local unknowns solved at each Gauss point, and the displacement~$\bm{u}$ and the temperature~$T$ are the global degrees of freedom. 

The indentation problem under consideration is approached using an axisymmetric finite-element formulation. Linear 4-noded quadrilateral elements are utilized for both the displacement $\bm{u}$ and the temperature $T$.
The computer implementation of the model is done by means of the automatic differentiation technique in AceGen, and the finite-element simulations are performed in AceFEM \citep{korelc2016}. Note that in the present simulations, the indenter is assumed to be rigid and frictionless. The corresponding contact problem is solved through enforcing the unilateral contact (impenetrability) condition on the potential contact surface using the augmented Lagrangian method~\citep{alart1991mixed}.

\subsection{Thermomechanical Analysis}\label{sec-analcoup}
The model described above is now applied to the problem of indentation of a polycrystalline NiTi. We analyze the temperature variation and its impact on the indentation-induced martensitic transformation across a broad range of spatial and temporal scales. Our primary objective is to identify the specific scales at which the transient thermal effects become substantial. Note that the spatial and temporal dependencies of the problem at hand can be readily recognized through the thermal diffusivity coefficient, which is defined as $k/(\varrho_0 c)$ and has the unit of square length per time~(m$^2$/s). 

We perform a series of simulations where the indenter radius $R$ is varied from 0.05~mm to 50~mm, thus covering a wide range of spatial scales from micro/nano-indentation to macro-indentation, and the indentation speed $v$ is varied from 0.1~$\mu$m/s to 100~$\mu$m/s, which is wide enough to encompass the range of practical speeds in quasi-static indentation tests, see e.g., \citep{amini2011loading,kumar2020assessment}. Note that the indenter radius of 50~mm exceeds the practical size of the indenter typically used in the experiment, and it is included here to provide a comprehensive assessment.

\begin{figure}
\centering
\includegraphics[width=0.5\textwidth]{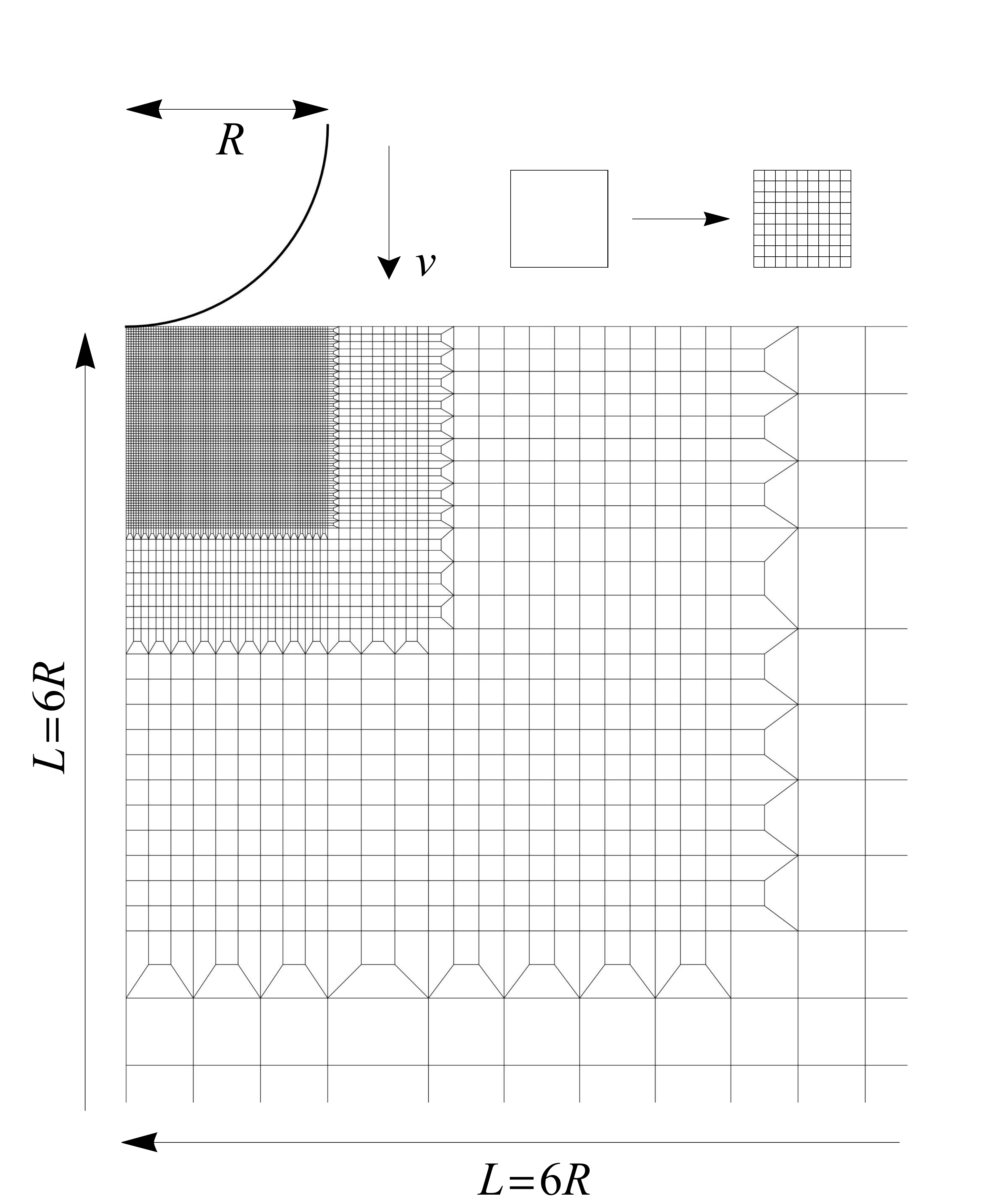}
\caption{Simulation setup for the axisymmetric indentation problem. The 2D mesh depicted here is 9 times coarser than the mesh actually used in the simulations, see the inset. To enhance the visibility of the mesh in the vicinity of the indenter, the entire computational domain is not displayed.}
\label{fig-mesh}
\end{figure}

The simulations are performed on a computational domain of the size $L \times L$, see Fig.~\ref{fig-mesh}. To preserve the geometrical similarity in all the simulations, the ratio of $L/R=6$ is kept fixed. Our preliminary analysis showed that the ratio $L/R=6$ is sufficiently large to avoid the spurious effects arising from the constrained boundaries. A non-uniform finite-element mesh is employed in which the area beneath the indenter possesses the finest mesh, with a ratio of $h/R=1.4 \times 10^{-3}$, with $h$ denoting the element size. The mesh gradually coarsens away from the indenter. In all the simulations, the loading is exerted up to a maximum normalized indentation depth of $\delta_\text{max}/R=0.04$. In experiments, larger indentation depths would likely induce considerable plastic deformation, thereby spoiling the pure pseudoelastic behavior.

The material parameters are set as follows: Young's modulus $E=41$~GPa, Poisson's ratio $\nu=0.3$, maximum transformation strain $\epsilon_\text{T}=0.039$ (within the typical range for NiTi in compression), specific entropy difference $\Delta s^*=0.24$~MPa/K, equilibrium temperature $T_\text{t}=255$~K, and critical driving force $f_\text{c}=4.6$~MPa. Parameters governing the heat transfer are taken from our previous studies \citep{RezaeeStupkiewicz2018Gradient,Rezaee2020Gradient}, namely specific heat capacity $\varrho_0 c=2.86$~MJ/(m$^3$ K) and heat conduction coefficient $k=18$~W/(m K). The initial temperature is set to $T_0=293$~K.

To facilitate meaningful comparisons of results across different spatial scales, size-dependent quantities should be normalized by suitable scaling factors. The normalization is adopted to eliminate trivial (first-order) geometrical effects, ensuring that any remaining variations in the results can be attributed to thermomechanical interactions. The indenter radius $R$ is a natural choice for normalizing the indentation depth~$\delta$. Regarding the indentation load $P$, an adequate normalization can be achieved by the nominal contact area, which not only removes the first-order geometrical effects but also yields a physically meaningful pressure-like quantity. We thus introduce the nominal contact radius $a^\text{nom}$, defined by simple geometry in terms of the indenter radius $R$ and the indentation depth $\delta$ as $a^\text{nom}=\sqrt{\delta (2R - \delta)}$. Specifically, $a^\text{nom}$ is the radius of a circle formed by the intersection of the indenter with the undeformed contact surface. The nominal contact area is then obtained by $A^\text{nom}=\pi (a^\text{nom})^2$. The value of $A^\text{nom}$ corresponding to the maximum indentation depth $\delta_\text{max}$, denoted as $A^\text{nom}_\text{max}$, is then used to normalize the load $P$. Conceptually, the nominal contact area is an idealized counterpart of the actual contact area that neglects the pile-up or sink-in deformations along the contact surface. Hence, it provides a scale-invariant representation of the load $P$ that is free from the surface deformation effects, and enables a straightforward comparison of the indentation responses. In Section~\nameref{sec-simpanal}, the nominal contact area is also utilized, as an alternative to actual contact area, for defining indentation hardness.

\begin{figure}
\hspace{-2.65cm}
\includegraphics[width=1.45\textwidth]{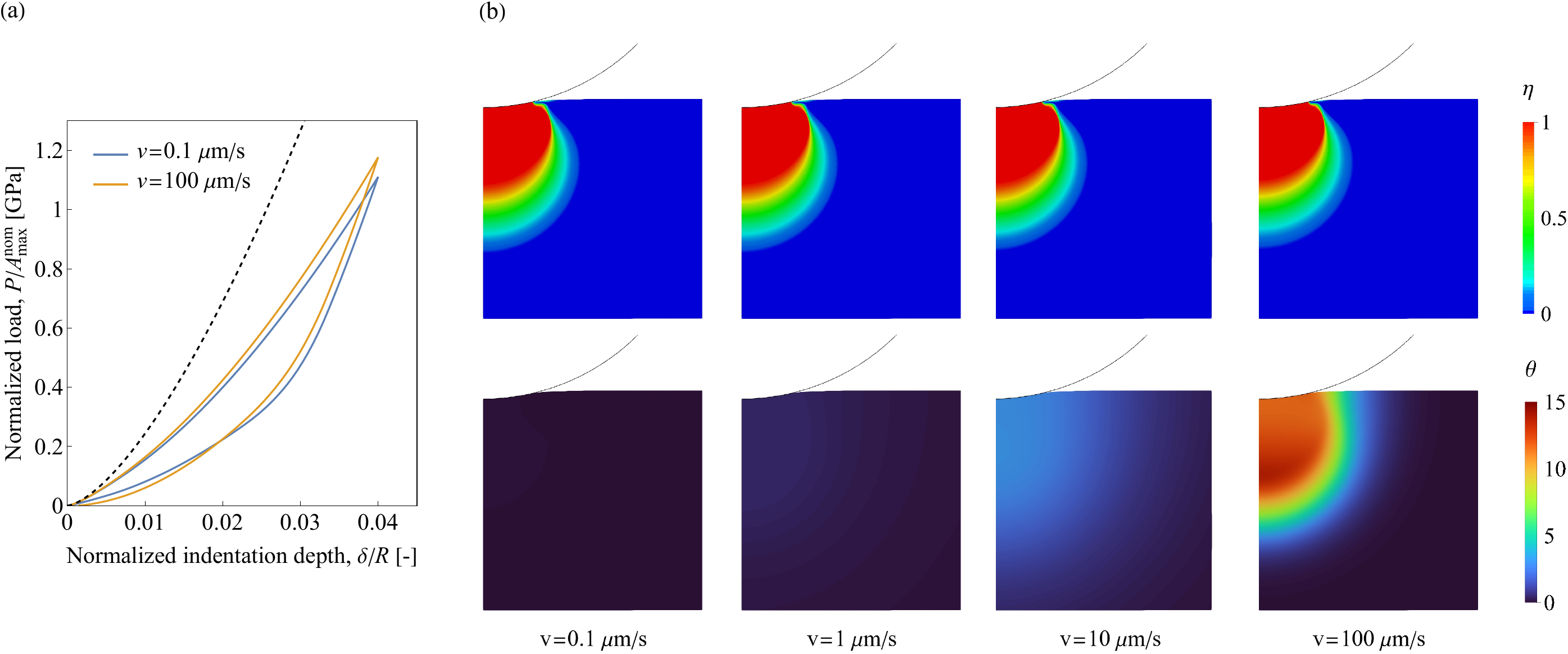}
\caption{Thermomechanically-coupled analysis of indentation-induced martensitic transformation: the effect of indentation speed $v$ on (a) normalized indentation load--depth response, and (b) spatial distribution of martensite volume fraction $\eta$ and relative temperature $\theta=T-T_0$. The results correspond to the case with the largest spatial scale, namely $R=50$ mm. The dashed curve in panel (a) shows the elastic response. The snapshots in panel (b) are taken at the maximum indentation depth of $\delta_\text{max}/R=0.04$.}
\label{fig-thermref}
\end{figure}

\begin{figure}
\centering
\includegraphics[width=1.0\textwidth]{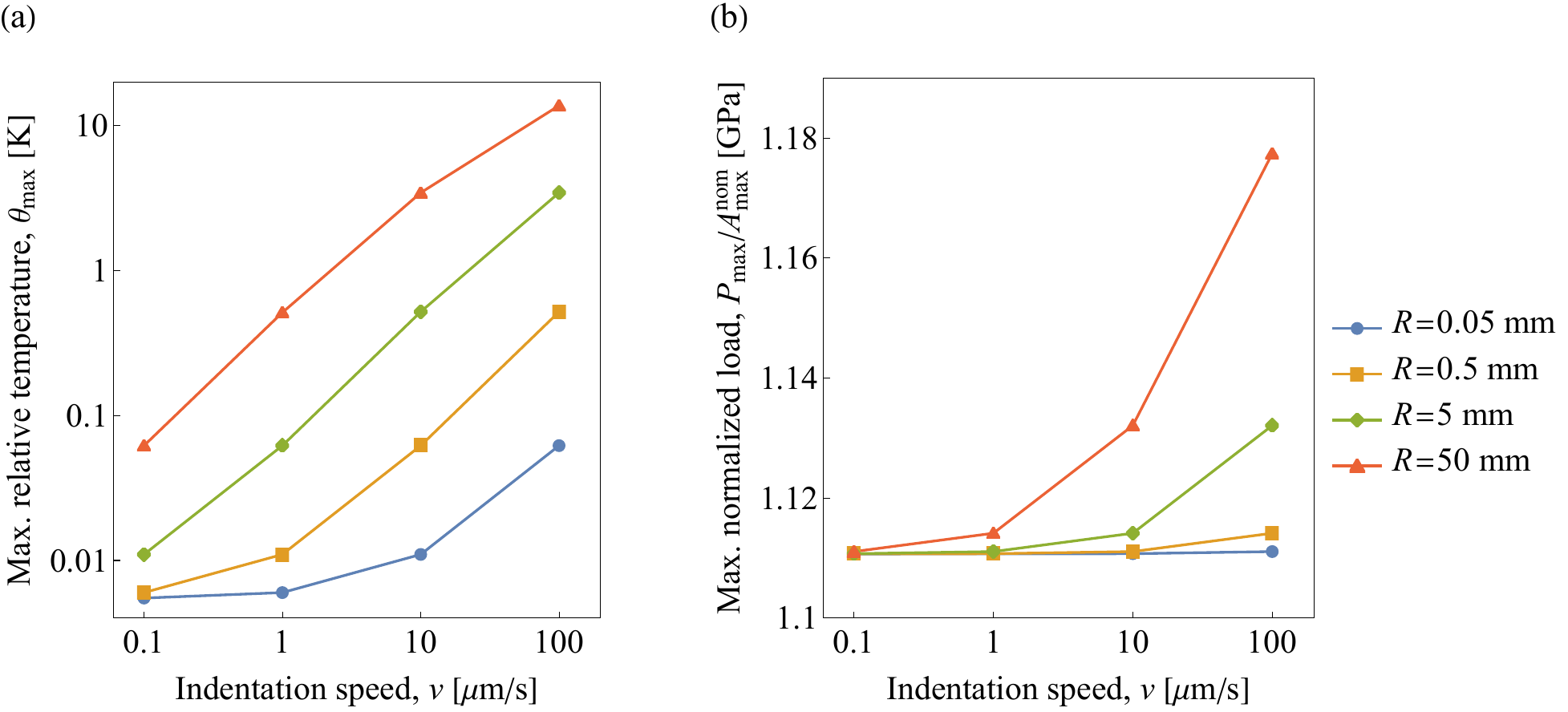}
\caption{Thermomechanically-coupled analysis of indentation-induced martensitic transformation: the maximum relative temperature $\theta_\text{max}$ (a) and the maximum normalized load $P_\text{max}/A_\text{max}^\text{nom}$ (b) as a function of the indentation speed $v$ for different spatial scales. The data correspond to the maximum indentation depth of $\delta_\text{max}/R=0.04$.}
\label{fig-thermmax}
\end{figure}

The simulation results indicate that at small spatial scales, the temperature distribution remains nearly uniform, with the relative temperature $\theta=T-T_0$ consistently below 1 K (recall that $T_0$ is the ambient temperature). This expected observation aligns with the dominant role of heat conduction at small spatial scales. As the spatial scale increases, $\theta$ becomes more pronounced, which then leads to the deviation of the indentation load--depth response from the isothermal case. This effect is illustrated in Fig.~\ref{fig-thermref} for the largest spatial scale with $R=50$ mm. It follows that at an indentation speed of $v=100$~$\mu$m/s, the normalized load at the maximum indentation depth exhibits an increase of about 66~MPa compared to the isothermal scenario, which stems from thermal hardening effects. Notably, during unloading, the indentation response follows a somewhat different trajectory, intersecting with the isothermal response at $\delta/R=0.02$. Interestingly, despite a maximum relative temperature of about $\theta_\text{max}=14$ K at $v=100$ $\mu$m/s, the distribution of martensite volume fraction $\eta$ seems to be unaffected and closely resembling that of the isothermal case. 

As a summary of this analysis, Fig.~\ref{fig-thermmax} depicts the variation of the maximum relative temperature $\theta_\text{max}$ and the maximum normalized load $P_\text{max}/A^\text{nom}_\text{max}$ with respect to the indentation speed $v$ for different spatial scales. It is important to note from Fig.~\ref{fig-thermmax} that the maximum relative temperature, reaching about $\theta_\text{max}=14$~K, remains visibly lower than $\theta_\text{max}=27.5$~K obtained under adiabatic condition (see Fig.~\ref{fig-simp}(a) in Section~\nameref{sec-simpanal}). This indicates that achieving adiabatic indentation would require much higher loading rates, higher than those feasible in quasi-static indentation experiments.

\subsection{Thermomechanical Analysis: the Role of TRIP Mechanism}\label{sec-trip}
Our investigation in the previous section was hinged on the assumption that pseudoelasticity is the sole inelastic mechanism governing deformation. This assumption is justified, given the use of spherical indenter and not targeting large indentation depths. Indeed, a handful of experiments have reported a complete recovery of inelastic deformation under shallow spherical indents~\citep{yan2006spherical,arciniegas2009thermoelastic,kumar2020assessment}. It is, however, important to note that in the majority of the micro/nano-indentation experiments, dislocation plasticity has been identified as an unavoidable contributor to inelastic deformation. In SMAs, plasticity typically arises from two main mechanisms: conventional yielding of the material under large external stresses, and the accumulation of dislocations and residual martensite that emit from microscopic phase transformation interfaces due to strain incompatibilities and large micro stresses. The latter mechanism, known macroscopically as transformation-induced plasticity (TRIP), is the primary cause of functional degradation in SMAs and is a subject of ongoing extensive research, see e.g., \citep{sidharth2024origins,vsittner2024recoverable,christison2025plasticity} for a few recent studies. Since our study adheres to shallow spherical indents, we disregard the involvement of conventional plastic yielding. Instead, we here investigate the extent to which the residual deformation of TRIP origin alters the observed thermal effects. Our model of pseudoelasticity is thus extended, in a simplified manner, to incorporate the TRIP mechanism. A brief discussion of this extension is provided in Appendix~\nameref{app-ext}. Below, we report the simulation results obtained with the extended model for the case with the largest spatial scale ($R=50$~mm) and largest indentation velocity ($v=100$~$\mu$m/s), i.e., the one showing the maximum thermal effects. While the large indenter radius and indentation velocity can be considered non-physical, we choose them to maximize the thermal effects, as already illustrated in the previous analysis of ideal pseudoelastic SMA.

\begin{figure}
\centering
\includegraphics[width=0.85\textwidth]{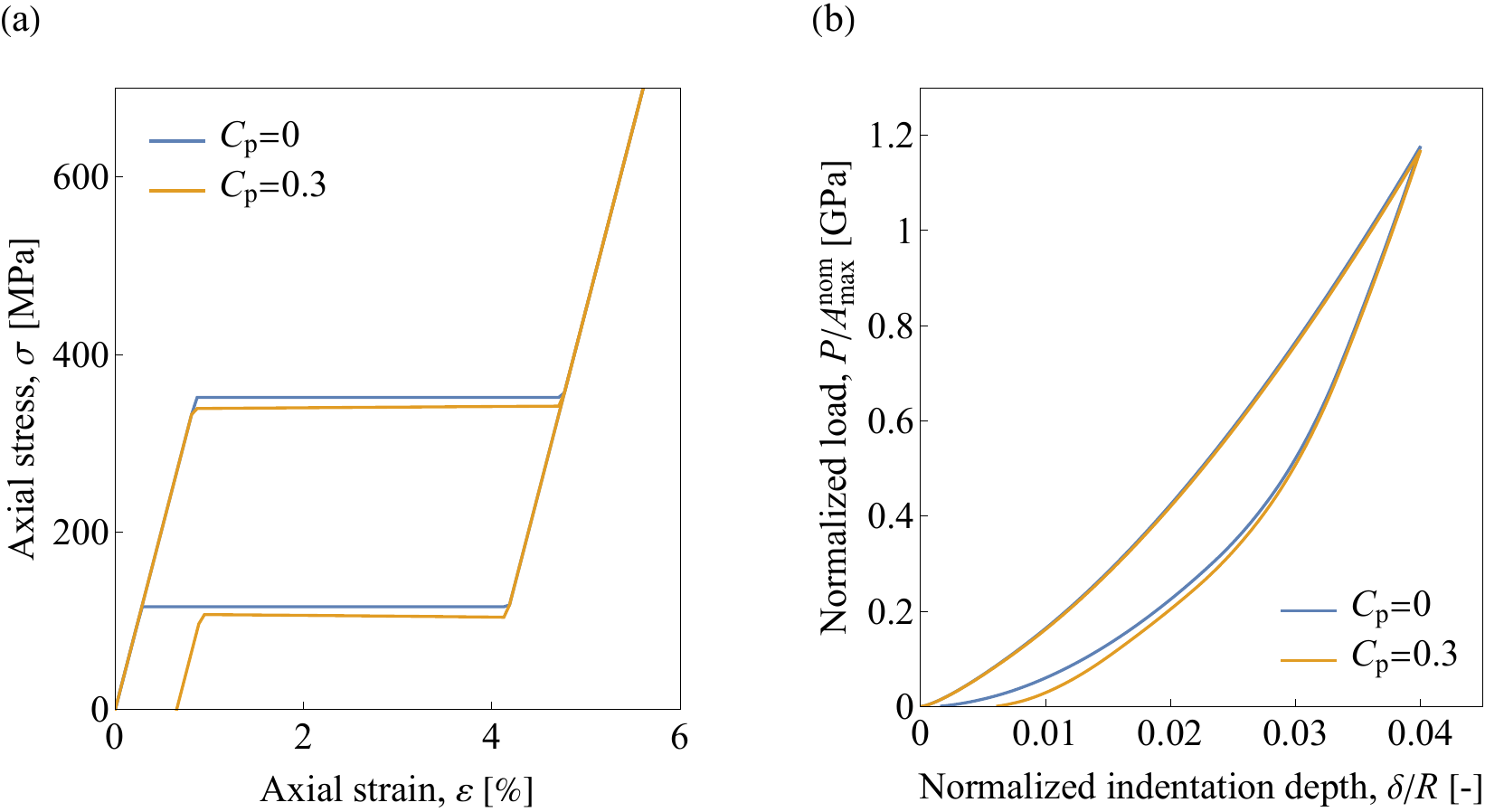}
\caption{Effect of TRIP mechanism on (a) material stress--strain response and (b) indentation load--depth response at an accumulation rate of $C_\text{p}=0.3$. The results in panel (b) correspond to the largest spatial scale and the largest indentation speed, specifically $R=50$~mm and $v=100$~$\mu$m/s.}
\label{fig-CpEffect}
\end{figure}

In the extended model, plasticity evolves through the accumulation of permanent strain $\bm{\varepsilon}^\text{p}$ and irreversible volume fraction $\eta^\text{ir}$, and is characterized by the respective saturation values, denoted as $\epsilon_\text{p}^\text{sat}$ and $h_\text{ir}^\text{sat}$, and the accumulation rate $C_\text{p}$. Following~\citep{rezaee2024modeling}, the saturation values are set at $40\%$ of their reversible counterpart limits, namely $\epsilon_\text{p}^\text{sat}=0.4 \epsilon_\text{T}$ and $h_\text{ir}^\text{sat}=0.4$. At the same time, we explored a broad range of $C_\text{p}$ to assess its influence on the results. It was revealed that only for moderate to large values of $C_\text{p}$ (where plasticity saturates within fewer than 50 full-transformation loading--unloading cycles) the indentation response does exhibit a noticeable deviation from the pseudoelastic one within the first cycle. Fig.~\ref{fig-CpEffect} compares the material stress--strain response and the normalized indentation load--depth response for $C_\text{p}=0.3$ with the corresponding pseudoelastic references ($C_\text{p}=0$). The material response shows a distinct buildup of residual strain, with an irreversible volume fraction of $\eta^\text{ir}=0.17$ and a permanent strain of $0.6\%$ accumulated within a single cycle. Analogously, as illustrated in Fig.~\ref{fig-CpEffect}(b), a discernible divergence of the indentation curves emerges during unloading once the reverse transformation initiates, and eventually, the TRIP mechanism results in a residual indentation depth of $\delta_\text{res}/R=0.006$, i.e., $15\%$ of the maximum indentation depth $\delta_\text{max}/R=0.04$.

\begin{figure}
\hspace{-0.3cm}
\includegraphics[width=1.05\textwidth]{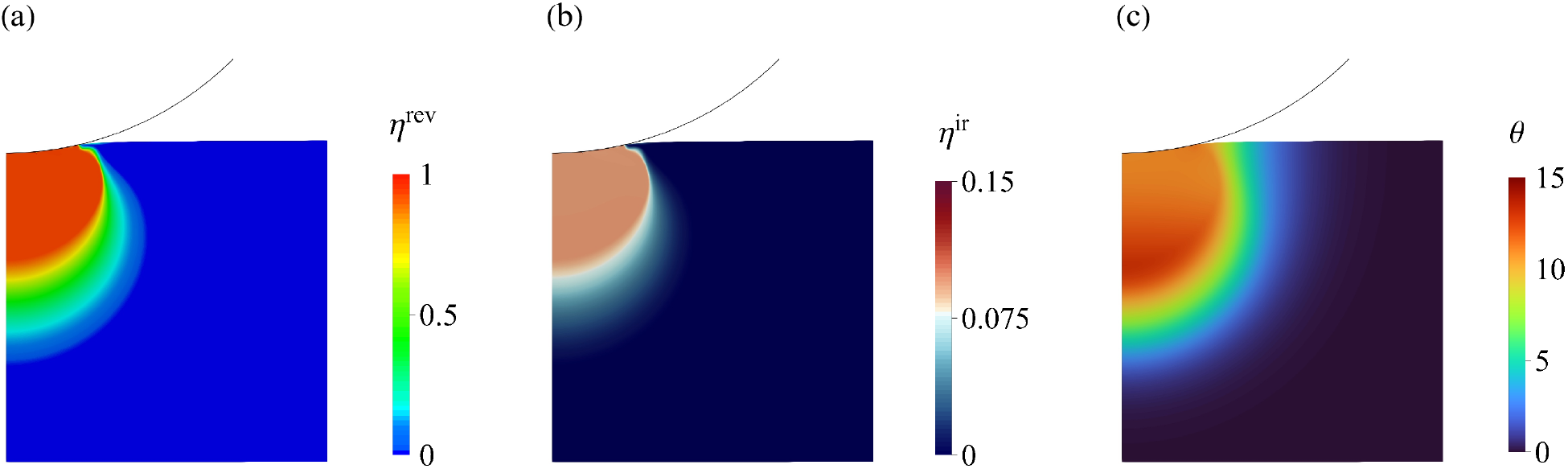}
\caption{Contribution of TRIP mechanism to the indentation-induced martensitic transformation at an accumulation rate of $C_\text{p}=0.3$: spatial distribution of (a) the reversible martensite volume fraction $\eta^\text{rev}$, (b) the irreversible martensite volume fraction $\eta^\text{ir}$, and (c) relative temperature $\theta=T-T_0$. The results correspond to the largest spatial scale and the largest indentation speed, specifically $R=50$~mm and $v=100$~$\mu$m/s.}
\label{fig-CpMic}
\end{figure}

\begin{figure}
\centering
\includegraphics[width=0.45\textwidth]{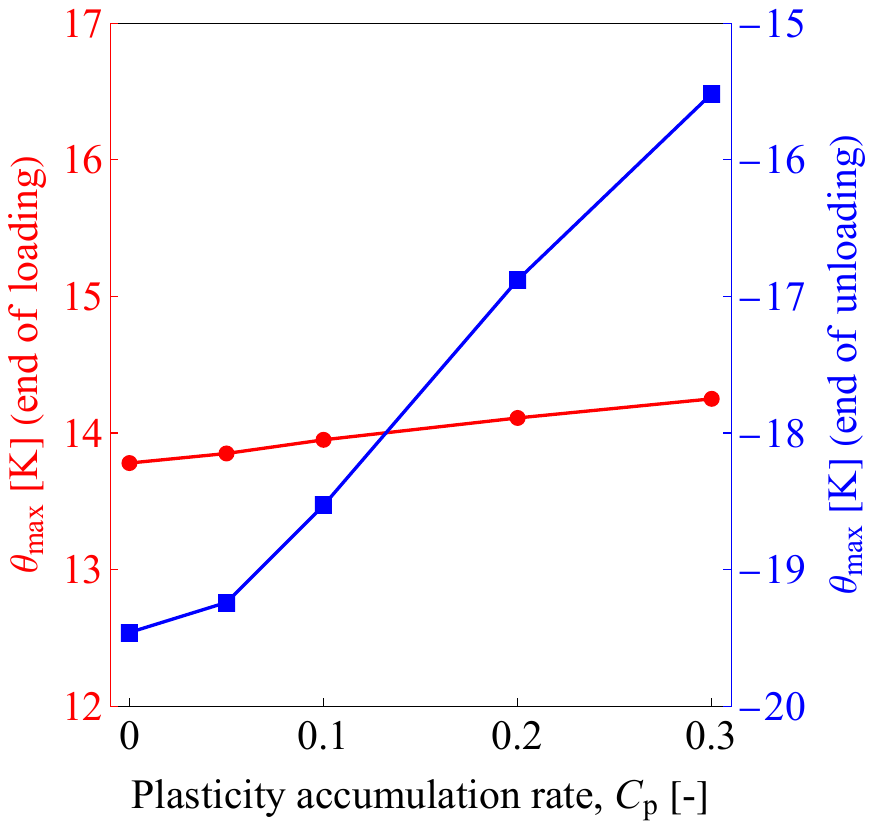}
\caption{Maximum relative temperature $\theta_\text{max}$ at the end of loading (red curve) and at the end of unloading (blue curve) as a function of plasticity accumulation rate $C_\text{p}$.}
\label{fig-CpEffectT}
\end{figure}

Fig.~\ref{fig-CpMic} presents the spatial distribution of the reversible martensite volume fraction $\eta^\text{rev}$, the irreversible volume fraction $\eta^\text{ir}$, and the relative temperature $\theta$ beneath the indenter at the end of loading. The distributions of $\eta^\text{rev}$ and $\theta$ resemble those observed in the ideal pseudoelastic case (see Fig.~\ref{fig-thermref}), albeit with somewhat different magnitudes. As a portion of the total martensite is converted into the irreversible one, the reversible volume fraction $\eta^\text{rev}$ exhibits a lower magnitude compared to the pseudoelastic case. The relative temperature $\theta$, however, shows a slight increase compared to the pseudoelastic case. As illustrated in Fig.~\ref{fig-CpEffectT}, this increase remains relatively minor. Even for the largest value of $C_\text{p}$ considered, the maximum relative temperature $\theta_\text{max}$ is only about 1~K higher than in the pseudoelastic case. This minor modification in thermal effects, as shown in Fig.~\ref{fig-CpEffect}(b), is too weak to exert an impact on the indentation response during loading and is overshadowed by the contribution of the TRIP mechanism, which itself remains small; the normalized indentation load at its peak is reduced by only 8 MPa, see Fig.~\ref{fig-CpEffect}(a). The result in Fig.~\ref{fig-CpEffectT} is extended to the end of unloading, showing that the maximum relative temperature $\theta_\text{max}$ exceeds the pseudoelastic reference by 4~K.

As the results show, even at the exaggerated large spatial scale and high indentation velocity examined, i.e., $R=50$~mm and $v=100$~$\mu$m/s, the influence of TRIP-induced residual deformation on the temperature change remains minimal. At smaller spatial scales and lower indentation velocities, this influence would be very weak, and effectively negligible. Finally, note that we refrain from considering higher values of $C_\text{p}$, as they induce one-cycle residual deformations significantly larger than those typically observed in uniaxial loading tests, see~\citep{rezaee2024modeling}.

\section{Thermal Effects during Adiabatic Indentation}\label{sec-adiabatic}
One of the key takeaways from the preceding analyses is that within the micro/nano spatial scales and at typical quasi-static indentation speeds, transient thermal effects and the resulting thermomechanical interactions are negligible. Nevertheless, in this section, with the aim to gain deeper insight into the thermomechanical aspects of the problem, a hypothetical scenario of adiabatic indentation is considered and the goal is to determine the potential extent to which thermal effects can contribute to the indentation behavior of SMAs. Given that the TRIP effects examined earlier are not substantial enough to reform the thermomechanical interactions, they are omitted from the present analysis. In what follows, we first describe the methodology used and then discuss the corresponding results.

\subsection{A Simplified Approach for Adiabatic Indentation}\label{sec-simp}
Our methodology follows the work of \citet{stupkiewicz2007micromechanics}, see Section~8.3.4 therein. It is based on the assumption that the adiabatic condition manifests solely through the increase of the transformation stress during the exothermic forward phase transformation, and analogously, through the decrease of the transformation stress during the endothermic reverse phase transformation. Accordingly, the model presented in Section~\nameref{sec-modelcoup} is simplified by treating the temperature $T$ as a dependent variable rather than an independent degree of freedom. The evolution of $T$ is then derived explicitly from the adiabatic heat balance equation, cf.~Eq.~\eqref{eq-heat} with $\bm{q}=\bm{0}$, and its variation in the chemical energy $\phi_\text{chem}$, see Eq.~\eqref{eq-chem}, leads to the increase/decrease of the transformation stress. 

During the forward transformation $(\dot{\eta}>0)$, the adiabatic heat balance equation can be expressed as the following differential equation (here, with $\eta$ playing the role of a pseudo-time)
\begin{equation}\label{eq-adfor}
    \varrho_0 c \frac{\rd T}{\rd \eta}=\Delta s^* T + f_\text{c}, \quad T(0)=T_0,
\end{equation}
and similarly during the reverse transformation $(\dot{\eta}<0)$ as
\begin{equation}\label{eq-adrev}
    \varrho_0 c \frac{\rd T}{\rd \eta}=\Delta s^* T - f_\text{c}, \quad T(1)=T_\text{m},
\end{equation}
where $T_\text{m}$ denotes the initial temperature of reverse transformation. The differential equations~\eqref{eq-adfor} and \eqref{eq-adrev} yield the following evolution equations for $T$ as a function of the martensite volume fraction $\eta$, respectively, 
\begin{equation}
    T-T_0=\left( T_0 + \frac{f_\text{c}}{\Delta s^*} \right)(e^{\eta \Delta s^*/\varrho_0 c}-1),
\end{equation}
and 
\begin{equation}
     T-T_\text{m}=\left( T_\text{m} - \frac{f_\text{c}}{\Delta s^*} \right)(e^{(\eta-1) \Delta s^*/\varrho_0 c}-1).
\end{equation}

\begin{figure}
\hspace{-2.6cm}
\includegraphics[width=1.3\textwidth]{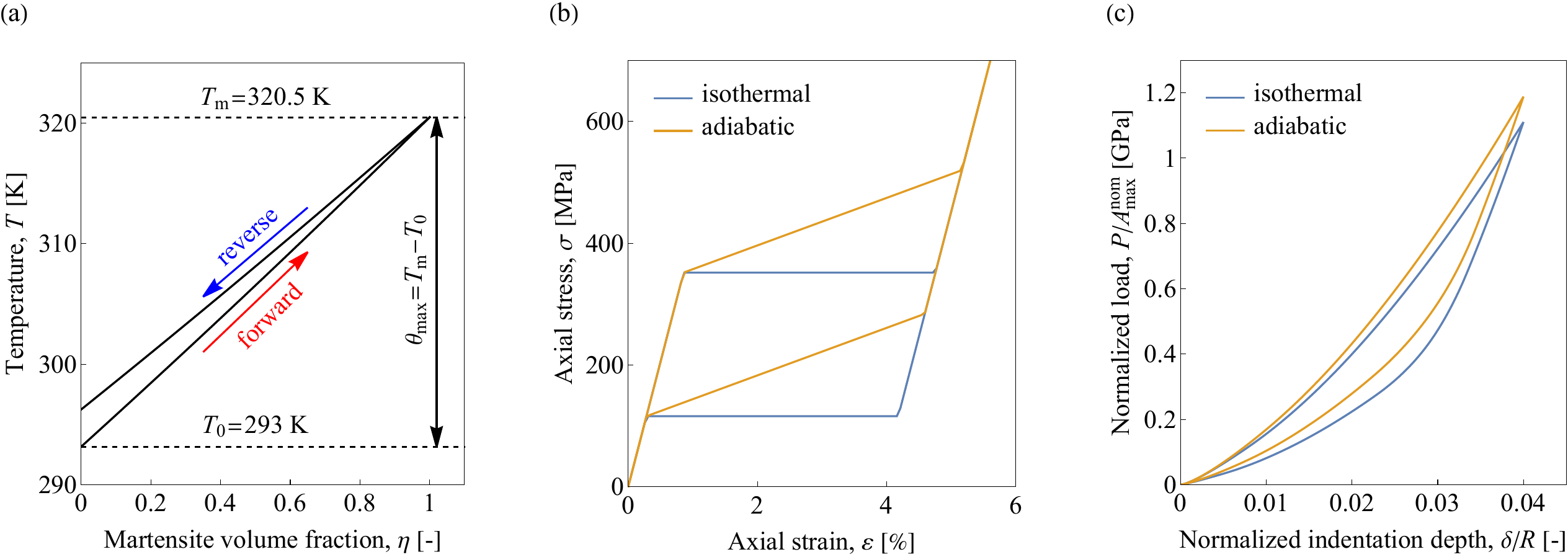}
\caption{A simplified procedure to mimic a (fully) adiabatic condition in martensitic transformation: (a) temperature change during the forward and reverse transformation, (b) the material response for the isothermal and adiabatic cases, and (c) the normalized indentation load--depth response for the isothermal and adiabatic cases. The calculations are performed for the initial temperature of $T_0=293$~K. The material parameters are the same as those introduced in Section~\nameref{sec-analcoup}.}
\label{fig-simp}
\end{figure}

Fig.~\ref{fig-simp}(a) depicts the evolution of temperature $T$ over a complete cycle of adiabatic forward and reverse transformation. In view of the contribution of the mechanical dissipation as a heat source, the temperature at the end of the cycle is marginally higher than the initial temperature $T_0$. In Fig.~\ref{fig-simp}(b), the adiabatic material response is compared with its isothermal counterpart. In the adiabatic case, the change in the chemical energy $\phi_\text{chem}$ induces a pronounced hardening behavior within the transformation regime. This, in turn, results in an indentation load--depth response with visibly higher loads compared to the isothermal case with a flat material response, as shown in Fig.~\ref{fig-simp}(c). In fact, this adiabatic response represents an upper-bound of the thermal effects in the indentation problem.

\subsection{Adiabatic Indentation vs.\ Isothermal Indentation}\label{sec-simpanal}
We carry out a comparative analysis to highlight the changes induced by the adiabatic condition with reference to the isothermal condition. Key quantities characterizing the indentation behavior are assessed, namely the hysteresis loop area $W^\text{hys}$, the volume of the transformed region $V^\text{tr}$, and the indentation hardness $H$. We investigate how these characteristic quantities vary with initial temperature $T_0$ across different indentation depths~$\delta$. Specifically, we consider the temperature range of $283 \leq T_0 \leq 323$ K (note that this parametric study is roughly equivalent to fixing the initial temperature $T_0$ and changing the equilibrium temperature $T_\text{t}$). The problem setup and the material parameters are the same as those in Section~\nameref{sec-analcoup}. Since the problem is now scale-independent, the findings remain the same regardless of the spatial scale. Nevertheless, to generalize the findings and to emphasize relative trends over magnitudes, the new quantities are reported in a normalized format. For this purpose, we introduce the reference volume $V^\text{nom}=2/3 \pi (a^\text{nom})^3$, representing the volume of a hemisphere with the nominal contact radius $a^\text{nom}$. The value of $V^\text{nom}$ corresponding to the maximum indentation depth $\delta_\text{max}$ is used as the scaling factor for the transformed volume $V^\text{tr}$. This normalization yields a scale-invariant measure that characterizes the extent of the transformed material relative to a hypothetical deformation zone beneath the indenter. Notice that in computing the hysteresis loop area $W^\text{hys}$ from the normalized indentation load--depth ($P/A_\text{max}^\text{nom}$--$\delta/R$) response, the scaling factor $R A_\text{max}^\text{nom}$ comes out naturally.

\begin{figure}
\centering
\includegraphics[width=0.88\textwidth]{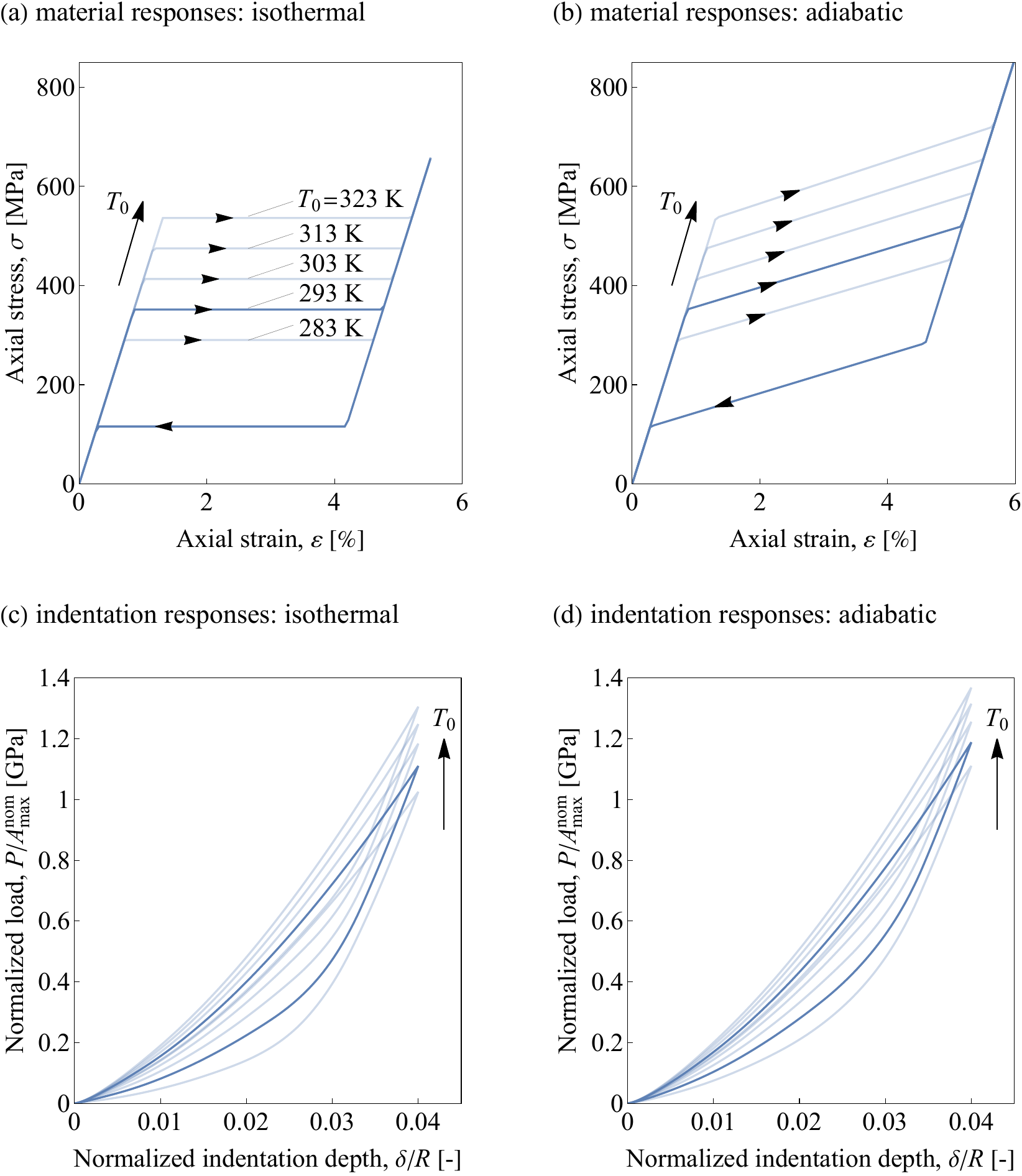}
\caption{The effect of the initial temperature $T_0$ on material response (a,b), and indentation load--depth response (c,d) in isothermal and adiabatic conditions. In panels (a,b), the full pseudoelastic loop is shown for $T_0=293$ K, while only the loading branches are shown for the other temperatures.}
\label{fig-ambient}
\end{figure}

\begin{figure}
\hspace{-2.7cm}
\includegraphics[width=1.45\textwidth]{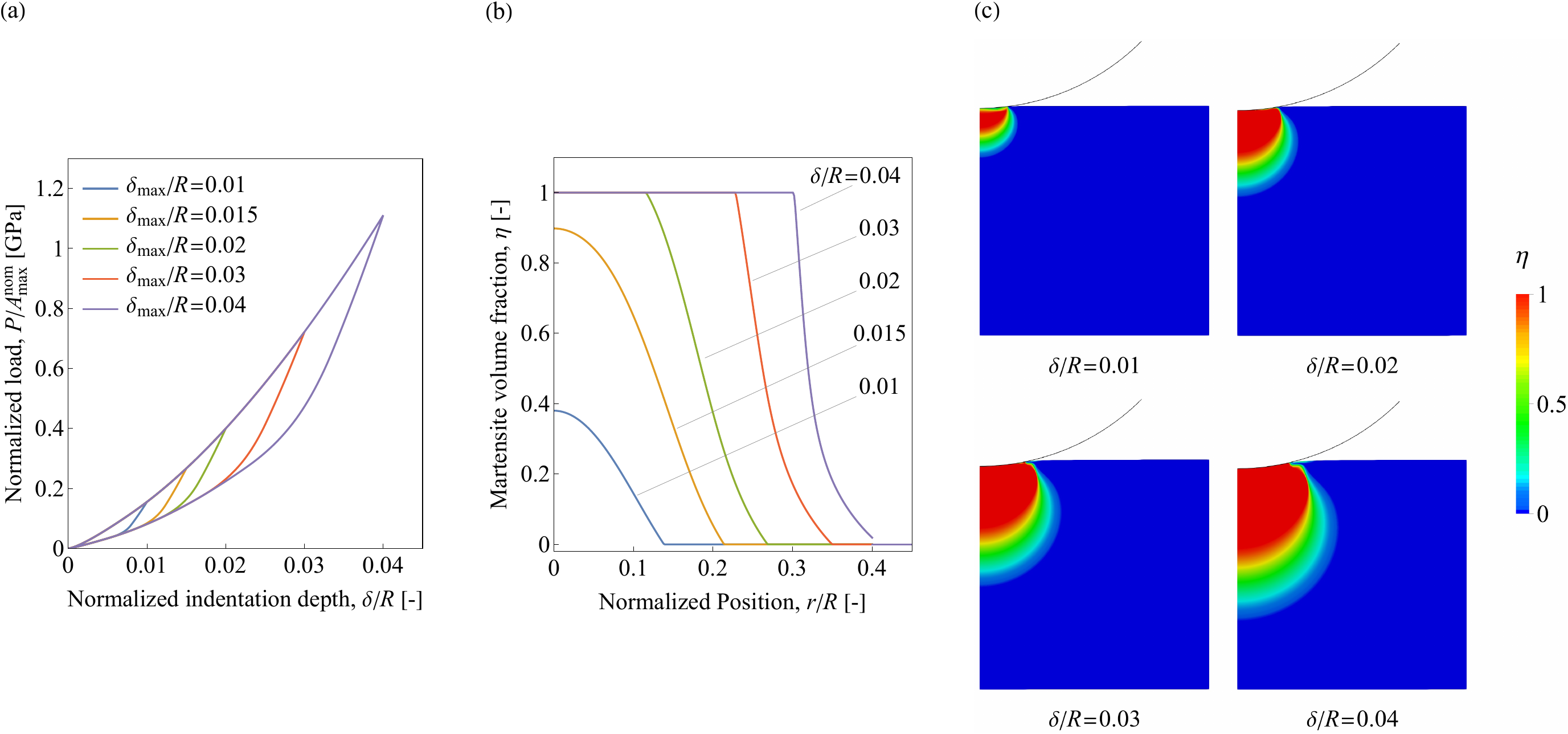}
\caption{Simulation results for the isothermal indentation at an initial temperature of $T_0=293$ K: (a) full-cycle normalized indentation load--depth response for various maximum indentation depths, and (b,c) profiles of the martensite volume fraction $\eta$ and snapshots of the transformed domain at different indentation depths. The profiles in panel (b) are taken at a normalized vertical position of $z/R=0.16$ beneath the indenter.}
\label{fig-demon}
\end{figure}

It is pertinent to begin the discussion by examining the effect of initial temperature $T_0$ on the material and indentation load--depth responses, as this facilitates the interpretation of the main analysis outcome presented subsequently. Fig.~\ref{fig-ambient} illustrates this effect for the isothermal and adiabatic scenarios. It follows from Fig.~\ref{fig-ambient}(a,b) that as $T_0$ increases, the transformation stress during the forward transformation (and obviously during the reverse transformation) increases. As a consequence, as can be seen from Fig.~\ref{fig-ambient}(c,d), the indentation loads increase with higher $T_0$. Additionally, Fig.~\ref{fig-ambient}(c,d) indicates that the hysteresis loop area $W^\text{hys}$ exhibits a noticeable variation as a function of $T_0$, a point that will be elaborated later on. Another important aspect to include in this discussion concerns the evolution of the (full-cycle) indentation load--depth response as a function of the maximum indentation depth $\delta_\text{max}$, as illustrated in Fig.~\ref{fig-demon}(a). A notable observation from Fig.~\ref{fig-demon}(a) is that, irrespective of $\delta_\text{max}$, the unloading branch consistently traces the same path. This suggests that the hysteresis loop area $W^\text{hys}$ likely exhibits a straightforward relationship with $\delta_\text{max}$. Finally, to complement the initial observations, Fig.~\ref{fig-demon}(b,c) showcases the progression of the transformed region at different indentation depths $\delta$. At each stage of transformation, the transformed region is divided into two distinct zones: a fully-transformed zone with martensite volume fraction of $\eta=1$ and a partially-transformed zone with $0 < \eta < 1$. It is evident from Fig.~\ref{fig-demon}(c) that as the indentation depth $\delta$ increases, not only the fully-transformed zone but also the partial one grows in size. The results presented in Fig.~\ref{fig-demon} correspond to the isothermal scenario, however, similar observations are applicable to the adiabatic scenario.

\begin{figure}
\centering
\includegraphics[width=0.92\textwidth]{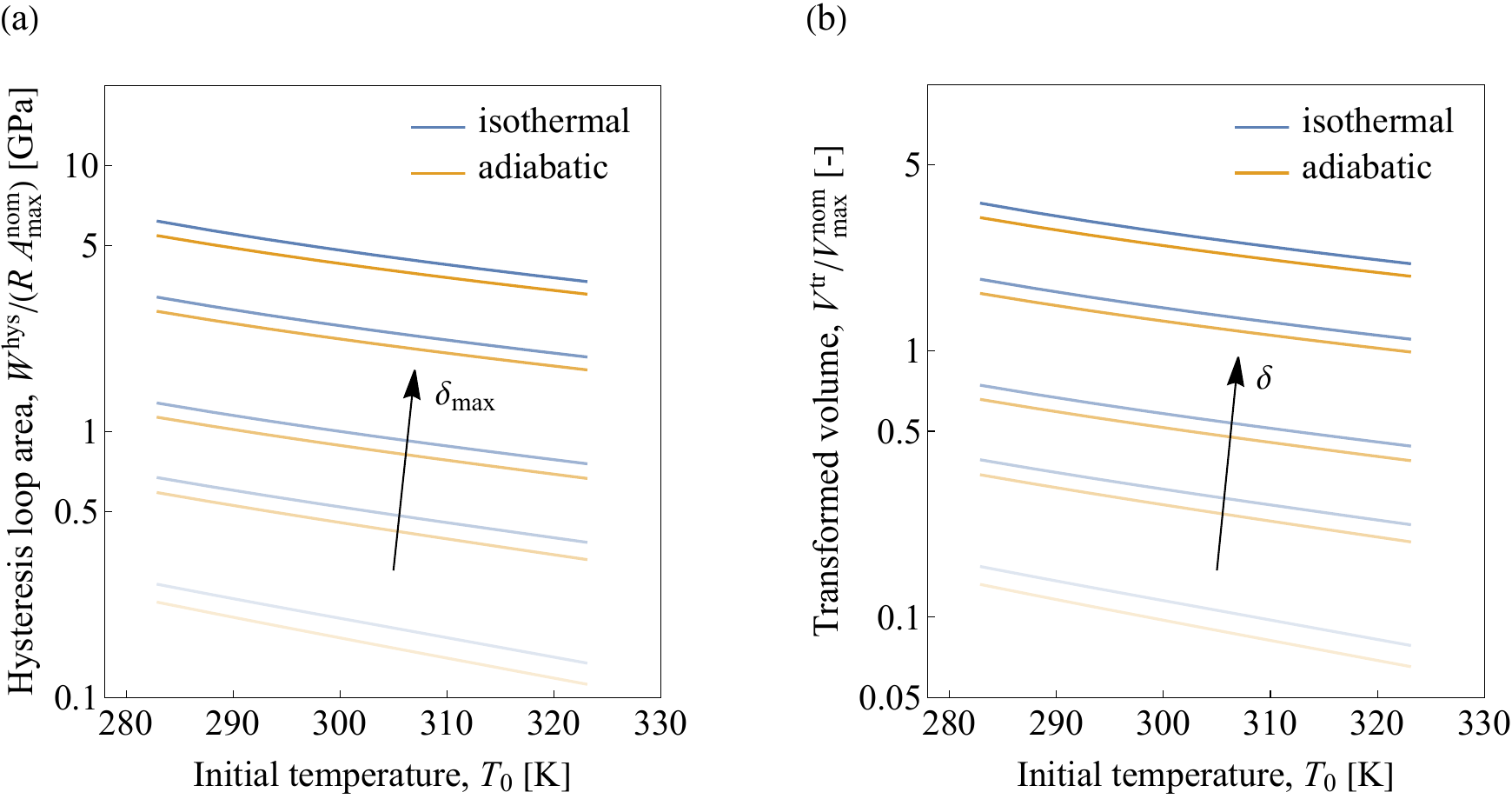}
\caption{Hysteresis loop area $W^\text{hys}$ (a) and volume of the transformed region $V^\text{tr}$ (b), both adequately normalized, as a function of the initial temperature $T_0$ and the indentation depth. Note the logarithmic--linear scale in both plots.}
\label{fig-hysvol}
\end{figure}

We now turn to the discussion of the characteristic quantities introduced earlier. Fig.~\ref{fig-hysvol} summarizes the overall behavior of the hysteresis loop area $W^\text{hys}$ and the volume of the transformed region $V^\text{tr}$, with the latter defined as the volume integral of the martensite volume fraction $\eta$. As shown in Fig.~\ref{fig-hysvol}, an identical trend is visible for both $W^\text{hys}$ and $V^\text{tr}$ (this is because the hysteresis loop area $W^\text{hys}$, i.e., dissipation, is related to the transformed volume $V^\text{tr}$ through the relation $W^\text{hys}=2 f_\text{c} V^\text{tr}$). More specifically, they both decline with increasing initial temperature $T_0$. Also, the gap between the adiabatic and isothermal curves becomes more pronounced as the indentation depth $\delta$ ($\delta_\text{max}$ for $W^\text{hys}$) increases (note the logarithmic scale in the vertical axes). Finally, the isothermal curve always lies above the corresponding adiabatic one. The latter effect is obviously an outcome of the hardening-type behavior in the adiabatic scenario, which results in a reduced indentation-induced transformation compared to the isothermal scenario, and hence a reduction in both the hysteresis loop area $W^\text{hys}$ and the transformed volume $V^\text{tr}$. By similar reasoning, the observed decreasing trend of the individual curves with increasing $T_0$ is also attributed to the increase of the transformation stress with $T_0$ (the Clausius--Clapeyron relation), as shown in Fig.~\ref{fig-ambient}(a,b), which, in turn, leads to a reduced transformation. 

\begin{figure}
\hspace{-3.5cm}
\includegraphics[width=1.5\textwidth]{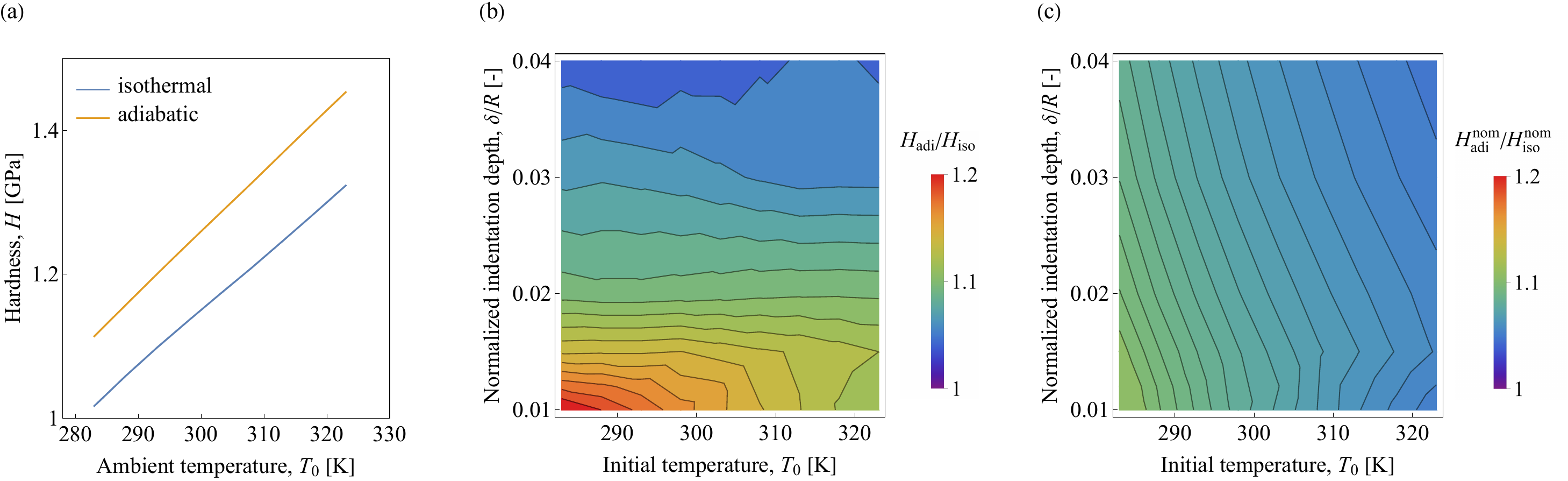}
\caption{Indentation hardness: (a) the individual adiabatic and isothermal curves for varying initial temperature $T_0$ at a normalized indentation depth of $\delta/R=0.02$, (b) contour plot of the hardness ratio $H_\text{adi}/H_\text{iso}$ as a function of the initial temperature $T_0$ and the normalized indentation depth $\delta/R$, and (c) contour plot of the nominal hardness ratio $H_\text{adi}^\text{nom}/H_\text{iso}^\text{nom}$, with the hardness values $H_\text{adi}^\text{nom}$ and $H_\text{iso}^\text{nom}$ calculated using the nominal contact area $A^\text{nom}$.}
\label{fig-hard}
\end{figure}

Finally, the results pertaining to the indentation hardness $H$ are discussed. The hardness $H$ is defined as
\begin{equation}
    H=\frac{P}{A},
\end{equation}
where $A$, the actual contact area, is obtained from the finite-element solution and corresponds to the same instant as the load $P$. As commented in Section~\nameref{sec-analcoup}, the finite-element mesh within the contact region has been significantly refined, ensuring that $A$, and thus hardness $H$, are computed with a reasonable accuracy. Fig.~\ref{fig-hard}(a) illustrates the relationship between the hardness $H$ and the initial temperature $T_0$ at a normalized indentation depth of ${\delta/R=0.02}$. Two immediate observations emerge from Fig.~\ref{fig-hard}(a): the adiabatic condition results in a higher hardness, and hardness increases with rising $T_0$. The effects are connected with the increase in the transformation stress, governed by the hardening-type behavior and the Clausius--Clapeyron relation, respectively. 

It is insightful to evaluate carefully the correlation between the adiabatic and isothermal hardness values (denoted, respectively, as $H_\text{adi}$ and $H_\text{iso}$). Fig.~\ref{fig-hard}(b) provides a contour plot of the hardness ratio $H_\text{adi}/H_\text{iso}$ over the entire range of indentation depth $\delta$ and initial temperature $T_0$ analyzed. The plot reveals that $H_\text{adi}/H_\text{iso}$ reaches its maximum at the smallest values of $\delta$ and $T_0$, while also exhibiting a steeper gradient within this region. Conversely, as $\delta$ and $T_0$ increase, $H_\text{adi}/H_\text{iso}$ approaches unity, and at the same time, its gradient diminishes. Concerning the influence of $T_0$, we note that the ratio between the total energy supplied to the system and the dissipated energy increases with increasing $T_0$ (as evident from the comparison between the area beneath the loading branch of the material response and the corresponding hysteresis loop area in Fig.~\ref{fig-ambient}(a,b)). As a result, at higher $T_0$, the type of the response within the transformation regime, hardening-type or flat, makes relatively smaller contribution to the indentation response, and this leads to a closer alignment between the adiabatic and isothermal loads. Concerning the influence of $\delta$, we note that a larger volume of material undergoes complete transformation at higher $\delta$ and the corresponding material response enters the stiff elastic branch of martensite, within which the adiabatic and isothermal cases behave the same. 

It is important to remark on the challenge of determining the contact area in experiments. In standard elasto-plastic materials, where unloading primarily involves elastic recovery, the contact area is either measured directly from the residual imprint or estimated analytically following the Oliver and Pharr method \citep{oliver1992improved}. In pseudoelastic SMAs, however, the reverse phase transformation that takes place during unloading causes the martensitic microstructure to recover (fully or partially), resulting in a residual imprint (if any) that does not represent genuinely the actual contact area under the maximum load \citep{liu1999indentation}, see also \citep{rezaee2023indentation} for a more detailed discussion. Accordingly, since the conventional methodologies used for elasto-plastic materials are not applicable in this context, an alternative approach is to use the nominal contact area~$A^\text{nom}$. In fact, the nominal contact area is also adopted for elasto-plastic materials, as a means to sidestep the experimental error resulting from the direct measurement of the contact area from the residual imprint \citep{petryk2017direct}.

Fig.~\ref{fig-hard}(c) presents the contour plot of the nominal hardness ratio $H_\text{adi}^\text{nom}/H_\text{iso}^\text{nom}$, with hardness values being evaluated using the nominal contact area $A^\text{nom}$. Evidently, when $A^\text{nom}$ is employed, a milder discrepancy is obtained between the adiabatic and isothermal hardness values. A comparison of the contour plots in Fig.~\ref{fig-hard}(b,c) reveals that the variation in the actual contact area $A$ does not follow the same trend as the indentation load $P$ (notice that the nominal hardness ratio $H_\text{adi}^\text{nom}/H_\text{iso}^\text{nom}$ actually represents the ratio between the corresponding indentation loads).
At lower indentation depths, $A$ in the adiabatic case is smaller than its isothermal counterpart, thus $A$ acts in concert with $P$ to yield a larger ratio $H_\text{adi}/H_\text{iso}$. At higher indentation depths, however, $A$ in the adiabatic case becomes larger than its isothermal counterpart, thus $A$ counteracts with $P$ to diminish the discrepancy between $H_\text{iso}$ and $H_\text{adi}$.

\section{Concluding remarks}\label{sec-sum}
In SMAs, thermomechanical interactions manifest in two primary aspects: the effect of ambient temperature, governed by the Clausius--Clapeyron relation, and the transient effect of latent heat of martensitic transformation. Both aspects are examined in the present study. Our analysis in Section~\nameref{sec-analcoup} demonstrates that thermomechanical coupling arising from the latent heat of transformation is of limited significance within the spatial and temporal scales relevant to quasi-static micro/nano-indentation problems. The results indicate that due to the predominant role of heat conduction, temperature gradients remain negligible, resulting in only a marginal influence on the indentation response. On the other hand, the analysis in Section~\nameref{sec-simpanal}, while underscoring the significant role of initial (ambient) temperature, reveals that even in the (hypothetical) adiabatic scenario, the simulation results do not show a substantial qualitative deviation from those in the isothermal scenario, thereby further highlighting the minor role of latent heat of transformation during indentation. This conclusion is supported by the observation that the volume of the transformed region $V^\text{tr}$, and likewise the hysteresis loop area $W^\text{hys}$, exhibit trends qualitatively similar in adiabatic and isothermal conditions, and also the related quantitative differences are not too high. Moreover, concerning the indentation hardness, the most significant variations occur at very shallow indents (and at low initial temperatures), a range that, given the limited volume of phase transformation, may not be the primary focus of interest in actual applications. By and large, our study suggests that incorporating thermal effects arising from the latent heat is not of critical importance when analyzing indentation-induced martensitic transformation, especially at small spatial scales. The complementary analysis in Section~\nameref{sec-trip} demonstrates that the residual deformations induced by TRIP mechanism weakly influence the thermomechanical coupling effects, thereby confirming the validity of our methodology and simulation results based on the ideal pseudoelasticity.

Since the main focus of the present study is on thermal effects, we have significantly simplified the SMA constitutive description. We have assumed an isotropic material with no tension--compression asymmetry and adopted a flat trilinear intrinsic material response. However, in reality, polycrystalline (textured) NiTi typically displays a strong transverse isotropy with a visible tension--compression asymmetry. Also, the material possesses a softening-type material response in tension and a hardening-type response in compression \citep{hallai2013underlying,reedlunn2014tension}. Nevertheless, we anticipate that our findings on thermomechanical interactions remain qualitatively the same regardless of the underlying constitutive relations adopted. It is also important to note that the softening-type behavior in NiTi tends to promote strain localization in thin specimens under tensile loads. However, since indentation is primarily associated with compressive stresses and involves bulk materials, the localization effects are expected to have limited relevance.

The present study is limited to a single loading--unloading indentation cycle. Under cyclic indentation, the combining effects of cyclic heat release/absorption, accumulation of residual strain and retained martensite, and the degradation of functional properties introduce additional complexities into the thermomechanical interactions. Nevertheless, the very limited temperature change observed in micro/nano-indentation simulations suggest that even when these factors are taken into account thermal effects are unlikely to play a major role in small-scale cyclic indentation. It is also important to point out that the conclusions drawn from our analysis pertain specifically to shallow spherical indents. In scenarios involving deep indents or sharp indenter geometries, conventional plastic yielding of austenite becomes a dominant inelastic mechanism that can have a significant impact on the thermomechanical interactions, and therefore cannot be simply neglected in the analysis.

\backmatter

\bmhead{Acknowledgements}
This work has been supported by the National Science Centre (NCN) in Poland through the Grant No.\ 2021/43/D/ST8/02555. For the purpose of Open Access, the authors have applied a CC-BY public copyright license to any Author Accepted Manuscript (AAM) version arising from this submission.

\begin{appendices}
\setcounter{equation}{15}

\section{Appendix: Model Extension to Incorporate TRIP Mechanism}\label{app-ext}
In this appendix, the model extension to include TRIP mechanism is briefly described. The formulation is based on the functional fatigue model recently developed in~\citep{rezaee2024modeling}. For simplicity, functional degradation effects that alter the material stress--strain response (such as the reduction of the transformation stress level, shrinkage of the hysteresis loop area, and the transition towards a hardening-type material response) are not considered in this extension. Two additional internal variables are included in the model to capture the TRIP effects. First, a permanent strain contribution, $\bm{\varepsilon}^\text{p}$, is introduced to represent plastic deformation. Second, an irreversible volume fraction, $\eta^\text{ir}$, is introduced that accounts for the retained (residual) martensite. Accordingly, the martensite volume fraction $\eta$ is split into the reversible part $\eta^\text{rev}$ and irreversible part $\eta^\text{ir}$. Thus, we have
\begin{equation}    \bm{\varepsilon}=\bm{\varepsilon}^\text{e}+\bm{\varepsilon}^\text{t}+\bm{\varepsilon}^\text{p}, \qquad \eta=\eta^\text{rev}+\eta^\text{ir}.
\end{equation}
Upon this extension, the transformation strain $\bm{\varepsilon}^\text{t}$ is redefined in terms of the reversible volume fraction, i.e., $\bm{\varepsilon}^\text{t}=\eta^\text{rev} \bar{\bm{\varepsilon}}^\text{t}$, and the inequality constraint on martensite volume fraction is modified to $0 \leq \eta^\text{rev} \leq 1- \eta^\text{ir}$, cf.~Eq.~\eqref{eq-const}. The extended model is based on the assumption that TRIP mechanism is active in parallel with martensitic transformation. Consequently, no separate yield surface or activation criterion is formulated for $\bm{\varepsilon}^\text{p}$ and $\eta^\text{ir}$. Instead, their evolution is linked to the martensitic transformation process via an accumulated volume fraction measure, $\eta^\text{acc}$, which is described as
\begin{equation}
    \dot{\eta}^\text{acc}=|\dot{\eta}^\text{rev}| \quad \Rightarrow \quad \eta^\text{acc}=\int_0^t |\dot{\eta}^\text{rev}| \rd \tau,
\end{equation}
and is governed by exponential-type evolution laws of the form
\begin{equation}\label{eq-tripev}
    \dot{\bm{\varepsilon}}^\text{p}= \epsilon_\text{p}^\text{sat} C_\text{p} \exp (-C_\text{p} \eta^\text{acc}) \dot{\eta}^\text{acc} \bm{N}_\text{p}, \quad \dot{\eta}^\text{ir}=h_\text{ir}^\text{sat}C_\text{p} \exp (-C_\text{p} \eta^\text{acc}) \dot{\eta}^\text{acc}.
\end{equation}
In Eq.~\eqref{eq-tripev}, $\epsilon_\text{p}^\text{sat}$ and $h_\text{ir}^\text{sat}$ denote the saturation values, $C_\text{p}$ characterizes the rate of plasticity accumulation, and the direction tensor $\bm{N}_\text{p}$ is defined as $\bm{N}_\text{p}=\bm{\varepsilon}^\text{t}/\lVert \bm{\varepsilon}^\text{t} \rVert$, with $\lVert \bm{\varepsilon}^\text{t} \rVert=\sqrt{\frac{2}{3} \bm{\varepsilon}^\text{t} \cdot \bm{\varepsilon}^\text{t}}$.

The Helmholtz free energy components take the same form as those in Eqs.~\eqref{eq-el} and \eqref{eq-chem}, with the chemical energy defined in terms of the total volume fraction $\eta=\eta^\text{rev}+\eta^\text{ir}$, while the dissipation potential $\Delta D$, cf.~Eq.~\eqref{eq-diss}, is now formulated in terms of the increment of the reversible volume fraction, namely $\Delta D =f_\text{c} |\Delta \eta^\text{rev}|$. Finally, the internal heat source, cf.~Eq.~\eqref{eq-heat} is extended to include the mechanical dissipation originating from the evolution of permanent strain and irreversible volume fraction, i.e.,
\begin{equation}
    \dot{R}=\Delta s^\ast T \dot{\eta}+f_\text{c} |\dot{\eta}^\text{rev}|+\bm{\mathcal{X}}_\text{p} \cdot \dot{\bm{\varepsilon}}^\text{p} + \mathcal{X}_\text{h} \dot{\eta}^\text{ir},
\end{equation}
with $\mathcal{X}_\text{h}=-\partial \phi/\partial \eta^\text{ir}$ and $\bm{\mathcal{X}}_\text{p}=-\partial \phi/\partial \bm{\varepsilon}^\text{p}$ as the respective thermodynamic driving forces. It is to be noted that the reversible martensite volume fraction $\eta^\text{rev}$ is the internal variable that is solved through the minimization problem, cf.~Eq.~\eqref{eq-min}, whereas the variables $\bm{\varepsilon}^\text{p}$ and $\eta^\text{ir}$ are explicitly determined based on the evolution equations~\eqref{eq-tripev}.

\end{appendices}

\bibliography{bibliography}%

\end{document}